\documentclass{article}

\usepackage{arxiv}

\usepackage{times}
\usepackage[ruled,vlined]{algorithm2e}
\usepackage{latexsym}
\usepackage{xurl}
\usepackage{xcolor}
\usepackage{colortbl}
\usepackage{multirow}
\usepackage{mathtools}
\usepackage[export]{adjustbox}
\usepackage{amsmath}

\usepackage{subfigure}

\usepackage{scalerel,stackengine}
\stackMath
\newcommand\reallywidehat[1]{%
\savestack{\tmpbox}{\stretchto{%
  \scaleto{%
    \scalerel*[\widthof{\ensuremath{#1}}]{\kern-.6pt\bigwedge\kern-.6pt}%
    {\rule[-\textheight/2]{1ex}{\textheight}}
  }{\textheight}%
}{0.5ex}}%
\stackon[1pt]{#1}{\tmpbox}%
}
\usepackage{bbm}

\usepackage{microtype}

\title{Empathy and Hope: Resource Transfer to Model Inter-country Social Media Dynamics}

\author{
Clay H. Yoo \\
  \small{Carnegie Mellon University} \\
  \texttt{hyungony@andrew.cmu.edu} \\
  \And
Shriphani Palakodety \\
  \small{Onai}\\
  \texttt{spalakod@onai.com} \\
 \And
Rupak Sarkar\\
  \small{Maulana Abul Kalam Azad University}\\
  \texttt{rupaksarkar.cs@gmail.com} \\
\And
Ashiqur R. KhudaBukhsh\thanks{Ashiqur R. KhudaBukhsh is the corresponding author.} \\
  \small{Carnegie Mellon University}\\
  \texttt{akhudabu@cs.cmu.edu} \\
}

\begin{document}
\maketitle

\begin{abstract}

The ongoing COVID-19 pandemic resulted in significant ramifications for international relations ranging from travel restrictions, global ceasefires, and international vaccine production and sharing agreements. Amidst a wave of infections in India that resulted in a systemic breakdown of healthcare infrastructure, a social welfare organization based in Pakistan offered to procure medical-grade oxygen to assist India - a nation which was involved in four wars with Pakistan in the past few decades. In this paper, we focus on Pakistani Twitter users' response to the ongoing healthcare crisis in India. While \#IndiaNeedsOxygen and \#PakistanStandsWithIndia featured among the top-trending hashtags in Pakistan, divisive hashtags such as \#EndiaSaySorryToKashmir simultaneously started trending. Against the backdrop of a contentious history including four wars, divisive content of this nature, especially when a country is facing an unprecedented healthcare crisis,  fuels further deterioration of relations. In this paper, we define a new task of detecting \emph{supportive} content and demonstrate that existing \emph{NLP for social impact} tools can be effectively harnessed for such tasks within a quick turnaround time. We also release the first publicly available data set\footnote{Data is publicly available at \url{https://github.com/anton-sturluson/empathy-and-hope}.} at the intersection of geopolitical relations and a raging pandemic in the context of India and Pakistan.

\end{abstract}


\section{Introduction}

The COVID-19 pandemic started in late 2019~\cite{carvalho2021first} and as of this writing is still ongoing. Several factors - geopolitical, economic, social among others - dramatically influenced health outcomes around the world. In this paper, we focus on the ongoing (as of May 2021) infection wave in India~\cite{CNN}. After aggressive initial steps to successfully curb the spread of the virus, case counts exploded in India towards the end of April 2021. The rapidity of the spread overwhelmed the healthcare infrastructure in the country. A widespread shortage of medical-grade oxygen~\cite{BBC}, overworked medical staff, and full capacity emergency rooms became the norm in major population centers.

\renewcommand{\tabcolsep}{2mm}
\begin{table*}[t]
{
\small
\begin{center}
     \begin{tabular}{|p{0.25\textwidth}|p{0.6\textwidth}|}
    
    \hline

\cellcolor{blue!15}\#PakistanStandsWithIndia & \cellcolor{blue!15} we're rivals not enemies. we breath same air speak same languages. our prayers , wishes and thoughts are with our brothers from other side of the border. We need to fight this bettle together \\
   \hline 
\cellcolor{red!15}\#PakistanstandswithIndia & \cellcolor{red!15} karma is bitch, india deserves what's happening right now because that's what they did with people of kashmir. kashmir's can't take revenge but god has his plans for redemption. \\
    \hline
\cellcolor{blue!15}\#IndiaNeedsOxygen & \cellcolor{blue!15} Despite the fact that we have our political conflicts, but I really pray for their good health. Get well soon india. Pakistani nation is with you. \\
    \hline
\cellcolor{red!15}\#IndiaNeedsOxygen & \cellcolor{red!15} India deserves this . You are facing what you did to kashmir and fool pakistani supporting india on this you are just slaves to british thats all .. \\
    \hline
\cellcolor{blue!15}\#EndiaSaySorryToKashmir & \cellcolor{blue!15}Kashmir is our and it is all of it. Until the independence of Kashmir, there will be war till the destruction of India. \\
    \hline
\cellcolor{red!15}\#EndiaSaySorryToKashmir &  \cellcolor{red!15}Political differences have their place but the prayers of us Pakistanis are with our Indian brothers and sisters.  May Allah give health to all. \\
\hline
    \end{tabular}
    
\end{center}
\caption{{Example tweets where the hashtag and the tweet content agree (highlighted in blue) and disagree (highlighted in red).}}
\label{tab:misalign}}
\end{table*}

The crisis was heavily discussed on social media and the associated hashtags were among the most discussed Twitter trends globally. In Pakistan, a neighboring country that fought four wars with India over the past seven decades~\cite{paul2005india}, a significant volume of tweets expressed solidarity with the Indian populace primarily through two hashtags - \#IndiaNeedsOxygen and \#PakistanStandsWithIndia. In addition, the hashtag \#EndiaSaySorryToKashmir started trending in Pakistan. The tweets using this hashtag were primarily divisive and often referenced a long-running territorial dispute at the heart of India-Pakistan relations. Amidst a far-reaching and rapidly progressing pandemic, divisive content of this nature negatively impacts the mental well-being of the affected population and can contribute to strained relations.


Hashtag based filtering, while extremely effective, cannot solely identify \emph{supportive} content. For instance, users can hijack trending hashtags and post content that violates the spirit of the hashtag (see Table~\ref{tab:misalign}). Also, replies or responses to a controversial tweet with a divisive hashtag may still retain the same hashtag but the content may reflect a unifying message. Rapidly evolving crises also require a fast turnaround time which can preclude sophisticated, time-consuming solutions.

In this paper, we present a method to automatically detect  \emph{supportive}  content from the tweet text (excluding hashtags, mentions, emojis, and urls). Our minimally supervised approach combines multiple soft signals - a \emph{hope speech} classifier that detects peace-seeking content~\cite{ECAIHopeSpeech}, and an \emph{empathy-distress} classifier trained on a well-known empathy-distress data set~\cite{buechel-etal-2018-modeling}. We further demonstrate superior performance in presence of supervision and release an annotated data set in this important humanitarian domain.  

Model reusability is a major challenge in NLP applications~\cite{arango2019hate,Beltagy2019SciBERT}.  We see our paper as preliminary evidence that NLP methods for positive impact research are not isolated efforts, and solutions arising from adjacent tasks can be re-purposed to tackle newer challenges.


\noindent\textbf{\emph{NLP for positive impact:}} Our work can be described by the following two broad themes specific to this workshop - \emph{online well-being \& positive information sharing} and \emph{case studies for NLP for social good}. In order to create a positive impact, we believe a research contribution needs to satisfy a subset of the following conditions: (1) a problem domain with a high societal impact; (2) resource-sharing to facilitate scientific progress; and (3) a research theme that spawns a rich line of follow-up work. 

Our paper has the following contributions:\\
\noindent\textbf{\emph{Social:}} We analyze the bilateral relationship between countries with a contentious history amidst a raging pandemic. Our work is at the intersection of two important themes - geopolitical relations and healthcare crises. We show a significant outpouring of support and solidarity between the two nations' online communities in the context of the pandemic. Barring a few recent efforts~\cite{ECAIHopeSpeech, TyagiSocInfo}, there is little literature on web manifestation of the India-Pakistan relationship co-occurring with other crises. To the best of our knowledge, this is the first analysis of social media text interactions between India and Pakistan amidst a pandemic. \\  
\noindent\textbf{\emph{Resource:}} We present a data set of tweets exploring geopolitical relations between historic adversaries amidst a health crisis. Publicly available data sets expressing empathy and distress are scarce~\cite{buechel-etal-2018-modeling}. Beyond our immediate objective of detecting \emph{supportive} tweets, this data set may be useful in answering several other research questions.\\ 
\noindent\textbf{\emph{The reusability argument:}} We present a compelling case study that \emph{NLP for positive impact applications} are not isolated tasks. Rather, multiple existing resources can be combined to tackle a new challenge in a fast turnaround time setting. 

\section{Task}

\renewcommand{\tabcolsep}{2mm}
\begin{table}[t]
{
\small
\begin{center}
     \begin{tabular}{|p{0.1\textwidth}|p{0.7\textwidth}|}
    
    \hline

Empathy &  Our hearts go out to our neighbours who are facing unprecedented misery. 

Pakistani People are praying for you \ldots \\
   \hline 
Distress &  I am a Pakistani but seriously this is heartbreaking what i am seeing from few days about India.We are enemies but this is about humanity,If we unite in this pandemic we both countries can fight together and can win this battle together,Peace \ldots\\
    \hline
Solidarity & As a human we all are together Pray for India and for all people all over the world who are suffering from COVID May Allah pak save us from this dangerous COVID-19 
Stop hating start praying  \\
    \hline

    \end{tabular}
    
\end{center}
\caption{{Example tweets exhibiting empathy, distress, and solidarity.}}
\label{tab:definition}}
\end{table}

In this paper, we consider the  task of detecting  \emph{supportive} content. Supportive behavior in language has been previously studied. For example, a AAAI-2020 shared task focused on detecting \emph{disclosure} and \emph{supportiveness} from  written accounts of casual and confessional conversations~\cite{chhaya2020editorial}. Our task is slightly different in the sense that we are interested in detecting content where speakers are supporting a country/people severely affected by a healthcare crisis.

We define \emph{supportive} content to be either expressing empathy, distress, or solidarity. Our definitions for empathy and distress follow~\cite{buechel-etal-2018-modeling} that considers extensive psychology literature~\cite{batson1987distress, batson1991evidence,sober1999unto,goetz2010compassion,mikulincer2010prosocial}. \cite{buechel-etal-2018-modeling} defines empathy as a warm, tender, and compassionate feeling for a suffering entity, and distress as a self-focused, negative affective
state that occurs when one feels upset due to
witnessing an entity’s suffering or need. Among the several existing definitions of solidarity, we borrow the following~\cite{wildt1999solidarity}:  a mutual
attachment between individuals (groups) that encompasses two levels: (1) a \emph{factual level} of
actual common ground between the individuals (groups); and (2) a \emph{normative level} of
mutual obligations to aid each other, as and when should be necessary. In Table~\ref{tab:definition} we present three example tweets exhibiting empathy, distress, and solidarity. 

Our definition for \emph{not-supportive} content does not have a similar psychological grounding. Our annotators observed that the \emph{not-supportive} content in this specific context, primarily  (1) expressed politically motivated hate;  (2) demonstrated a war-mongering attitude; (3) expressed schadenfreude; (4) mentioned politically contentious issues; and (5) expressed unrelated content such as product promotion etc.


\section{Resource}

We use two existing resources for our work. Next, we present a short description of these resources.  

\subsection{\emph{Hope speech} classifier}
The \emph{hope speech} detection task introduced in~\cite{ECAIHopeSpeech} involves identifying social media text content with a unifying message encouraging peace, discouraging war, and highlighting the economic, social, and human costs of conflict against the backdrop of the 2019 India-Pakistan conflict. A detailed definition of \emph{hope speech} with illustrative examples is provided in~\cite{ECAIHopeSpeech}. 


\subsection{Empathy and Distress Classifier} 

We train a classifier on the empathy-distress data set introduced in~\cite{buechel-etal-2018-modeling}. The data set is grounded in prior psychology literature on empathy and distress~\cite{batson1987distress, batson1991evidence,sober1999unto,goetz2010compassion,mikulincer2010prosocial}. The data set consists of 418 news article excerpts from popular news platforms and responses
to them from 403 annotators, resulting in a total of 2,015 responses (5 articles per annotator). Upon filtering 
responses that deviated from the task description,
the pruned final data set consists of 1,860 responses (empathy: 916, distress:
905). We split this data into train and test sets in 90/10 ratio and train a binary classifier using \texttt{BERT}~\cite{Devlin2019BERTPO} (\texttt{bert-base-uncased}) using transformers library~\cite{wolf-etal-2020-transformers}.

\section{Data}

Our data set, $\mathcal{T}$, consists of 309,394 tweets posted by 150,289 unique users collected between 21 April 2021 and 04 May 2021. The top trending hashtags in Pakistan for April 22 and April 23 were retrieved from \url{https://getdaytrends.com/} and all associated tweets were obtained using the Twitter API\footnote{https://developer.twitter.com/en/docs/twitter-api}. Other closely related trending hashtags were also included (e.g., \#IndiaNeedsOxygen and \#IndiaNeedOxygen, or \#PakistanStandsWithIndia and \#PakistanStandWithIndia). Additional details are in Table~\ref{tab:data_statistics}. In this paper, any mention of a hashtag includes closely spelled variants (e.g. \#IndiaNeed(s)Oxygen, \#PakistanStand(s)WithIndia, or \#I(E)ndiaSaySorryToKashmir). We define the following two hashtag sets: $\mathcal{H}_\mathit{supportive}$ = \small{$\{$\#IndiaNeed(s)Oxygen, \#PakistanStand(s)WithIndia$\}$} \normalsize; and $\mathcal{H}_{\mathit{not}\emph{-}\mathit{supportive}}$ = \small{$\{$\#I(E)ndiaSaySorryToKashmir$\}$}.\normalsize  

\noindent\textbf{Subsets of interest:} Two mutually disjoint subsets of $\mathcal{T}$: 
$\mathcal{T}_\mathit{supportive}$ and $\mathcal{T}_{\mathit{not}\emph{-}\mathit{supportive}}$ are defined as follows. $\mathcal{T}_\mathit{supportive}$ includes tweets containing one or more of the $\mathcal{H}_\mathit{supportive}$ hashtags and $\mathcal{T}_{\mathit{not}\emph{-}\mathit{supportive}}$ includes tweets containing one or more of the $\mathcal{H}_{\mathit{not}\emph{-}\mathit{supportive}}$ hashtags. Tweets containing any intersection of the $\mathcal{H}_\mathit{supportive}$ and $\mathcal{H}_{\mathit{not}\emph{-}\mathit{supportive}}$ hashtags are discarded from either subset and thus there is no intersection between  $\mathcal{T}_\mathit{supportive}$ and $\mathcal{T}_{\mathit{not}\emph{-}\mathit{supportive}}$. Since classification of extremely short texts is a well-established challenge~\cite{sindhwani2009uncertainty,attenberg2010unified,khudabukhsh2015building}, in all of our sampling experiments involving a text classifier, we impose a length restriction of 10 or more tokens after preprocessing. Furthermore, our classifiers are only presented with the tweet text, i.e., the body of the tweet with hashtags, emojis, urls, and mentions removed.   

\noindent\textbf{Generating country labels for tweets:} The Twitter API bundles geographic location (coordinates) with tweets. In addition, we utilized a weak signal - if a user's Twitter handle contains an India or Pakistan flag emoji, then we assume their tweets originated in India or Pakistan respectively. In the cases where the location information and our signal are both present, we notice no inconsistency, indicating our weak country signal is robust.

\section{Characterization of the Tweets} 

\subsection{Likes and Retweets}\label{sec:likeRetweet}

We now characterize the retweets and likes of each of these hashtags. Let \#$\mathit{ht}_\mathit{Ind}$, \#$\mathit{ht}_\mathit{Pak}$, and \#$\mathit{ht}_\mathit{Other}$ denote the subsets of tweets that contain the hashtag $\mathit{ht}$ and originate in India, Pakistan, and other (or unknown), respectively. Table~\ref{tab:data_like} shows that overall, the tweets containing \emph{supportive} hashtags received fewer likes and retweets than those containing \emph{not-supportive} hashtags. We further notice that tweets containing \emph{supportive} hashtags that originated in Pakistan received substantially more likes than those from India. Our results though come with the following caveats. Multiple factors can influence our data collection process such as the inner workings of Twitter algorithms or the Twitter API. Also, our focus is on English tweets; previous studies have reported that Hindi is more commonly used
to express negative sentiment in social media content generated in the Indian sub-continent~\cite{rudra-etal-2016-understanding,KhudaBukhshHarnessing}. 

\subsection{Hashtag Co-occurrence}

\begin{figure*}[htb]

\centering
\subfigure[\#IndiaNeed(s)Oxygen]{%
\includegraphics[frame,trim={1.5mm 2.5mm 1.5mm 2.5mm}, clip, width = 0.3 \textwidth]{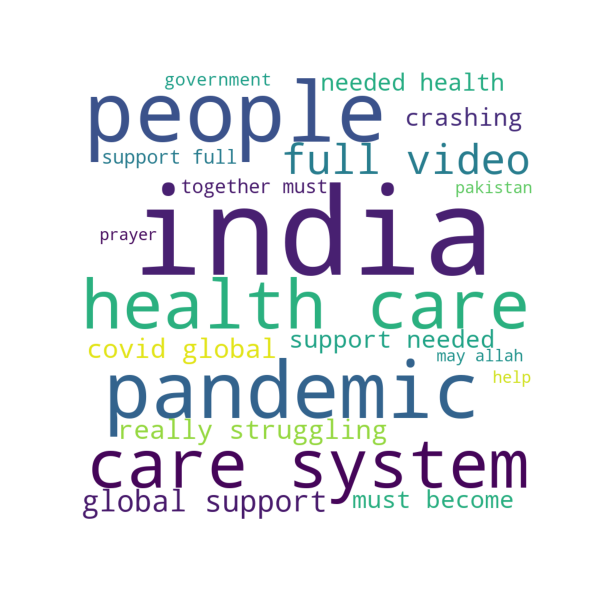}
\label{fig:Oxygen}}
\subfigure[\#PakistanStand(s)WithIndia]{%
\includegraphics[frame,trim={1.5mm 2.5mm 1.5mm 2.5mm}, clip, width = 0.3 \textwidth]{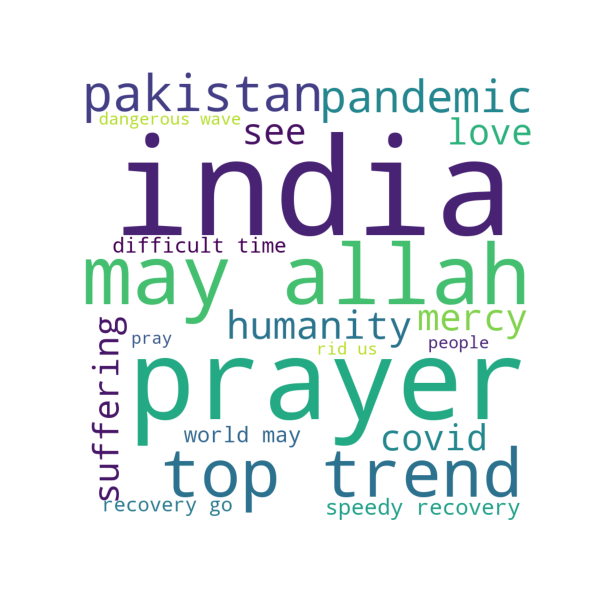}
\label{fig:PakistanStands}}
\subfigure[\#I(E)ndiaSaySorryToKashmir]{%
\includegraphics[frame,trim={1.5mm 2.5mm 1.5mm 2.5mm}, clip, width = 0.3 \textwidth]{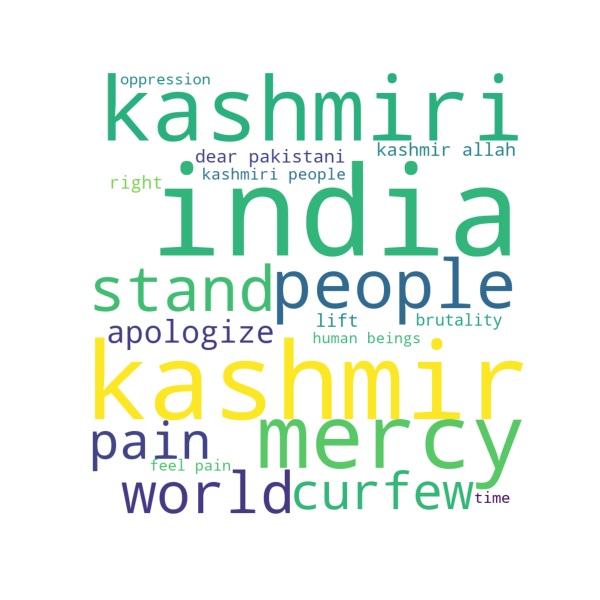}
\label{fig:EndiaSaySorry}}
\caption{\small{A word cloud visualization of the tweet contents and the associated hashtag used. Hashtags and punctuations are removed as a preprocessing step.}}
\label{fig:visualization}
\end{figure*}

We next measure in-group and out-group co-occurrence of \emph{supportive} and \emph{not-supportive} hashtags within a single tweet. Pair-wise Jaccard index between the tweet sets using various hashtags is computed\footnote{Jaccard index is a statistic to gauge similarity between two sets, $\mathcal{A}, \mathcal{B}$, expressed as $\frac{|\mathcal{A} \cap \mathcal{B}|}{|\mathcal{A} \cup \mathcal{B}|}$.} and shown in Table~\ref{tab:jaccard}. We observe that among all hashtag pairs, $\langle$\#IndiaNeed(s)Oxygen and \#PakistanStand(s)WithIndia$\rangle$ occurs the most. We observe that qualitatively, there is a stark contrast between tweets containing $\mathcal{H}_\mathit{supportive}$ hashtags and tweets containing $\mathcal{H}_{\mathit{not}\emph{-}\mathit{supportive}}$ hashtags with the dominant theme in the former being empathy, distress, and solidarity. Figure~\ref{fig:visualization} presents a word-cloud visualization of the tweets employing the three hashtags. 

\begin{table*}[htb]
\resizebox*{\textwidth}{!}{%
\begin{tabular}{l|lll}
hashtags     & \#IndiaNeed(s)Oxygen & \#PakistanStand(s)WithIndia & \#I(E)ndiaSaySorryToKashmir   \\ \hline
\#IndiaNeed(s)Oxygen     & -      & 0.0887 & 0.0247 \\
\#PakistanStand(s)WithIndia     & 0.0887 & -      & 0.0405 \\
\#I(E)ndiaSaySorryToKashmir     & 0.0247 & 0.0405 & -     
\end{tabular}
}
\vspace{0.5cm}
\caption{Jaccard index of tweet subsets employing various hashtags.}
\label{tab:jaccard}
\end{table*}

\begin{table}[htb]
\centering
\resizebox*{0.60\textwidth}{!}{%
\begin{tabular}{llll}
\hline
Hashtag & Total   & India  & Pakistan \\ \hline
\#IndiaNeedsOxygen            & 145,975 & 26,383 & 19,748   \\
\#IndiaNeedOxygen             & 24,488  & 5,049  & 2,400    \\
\#PakistanStandsWithIndia     & 96,226  & 12,331 & 21,583   \\
\#PakistanStandWithIndia      & 17,406  & 2,772  & 3,790    \\
\#EndiaSaySorryToKashmir      & 25,081  & 87     & 8,022    \\
\#IndiaSaySorryToKashmir      & 557     & 15     & 169      \\ \hline
All                           & 309,733 & 46,651 & 55,712 \\ \hline
\end{tabular}%
}
\vspace{0.5cm}
\caption{Statistics of dataset crawled between 21 April 2021 and 04 May 2021.}
\label{tab:data_statistics}
\end{table}

\begin{table*}[htb]
\small
\centering
\begin{tabular}{llll}
\hline
Hashtag$_{Location}$ & Like   & Retweet \\ \hline
\#IndiaNeed(s)Oxygen$_\mathit{Ind}$            & $2.32 \pm 63.80$  & $1631.07 \pm 3393.86$    \\
\#IndiaNeed(s)Oxygen$_\mathit{Pak}$            & $4.39 \pm 96.50$ & $322.98 \pm 1107.28$    \\

\#IndiaNeed(s)Oxygen$_\mathit{Other}$            & $2.72 \pm 215.81$ & $1306.71 \pm 2934.60$   \\
\hline 
\#PakistanStand(s)WithIndia$_\mathit{Ind}$            &  $2.46 \pm 78.14$ & $2313.45 \pm 2898.67$   \\
\#PakistanStand(s)WithIndia$_\mathit{Pak}$            &  $8.58 \pm 358.16$ & $665.03 \pm 1559.59$   \\
\#PakistanStand(s)WithIndia$_\mathit{Other}$            & $2.65 \pm 117.25$ & $1246.58 \pm 2195.85$   \\
\hline
\#I(E)ndiaSaySorryToKashmir$_\mathit{Ind}$            & $1.49 \pm 4.97$ & $191.45 \pm 266.38$  \\
\#I(E)ndiaSaySorryToKashmir$_\mathit{Pak}$            & $1.26 \pm 24.80$ & $276.28 \pm 300.87$   \\
\#I(E)ndiaSaySorryToKashmir$_\mathit{Other}$            & $1.51 \pm 37.61$ & $248.33 \pm 293.02$   \\
\hline
\end{tabular}%
\caption{Location-specific like and retweet behavior.}
\label{tab:data_like}
\end{table*}

\begin{table}[htb]
\centering
\resizebox*{0.60\textwidth}{!}{%
\begin{tabular}{llll}
\hline
Model & Precision & Recall   & F1 \\ 
\hline 
$\mathcal{M}^{\texttt{BERT}}_\mathit{supervised}$     & $83.28 \pm 0.8$  & $80.98 \pm 1.6$ & $81.14 \pm 1.6$ \\
$\mathcal{M}^{\texttt{BERT}}_\mathit{informed}$     & $80.78 \pm 0.5$ & $80.60 \pm 0.7$ & $80.62 \pm 0.6$ \\
$\mathcal{M}^{\texttt{BERT}}_
\mathit{hashtag}$       & $72.93 \pm 1.2$ & $53.78 \pm 1.6$ & $48.58 \pm 2.5$ \\
$\mathcal{M}^{\texttt{SVM}}_\mathit{supervised}$     & $66.38 \pm 0.5$ & $91.65 \pm 0.8$ & $76.99 \pm 0.4$ \\
$\mathcal{M}^{\texttt{SVM}}_\mathit{informed}$     & $56.98 \pm 0.6$ & $94.03 \pm 0.3$ & $70.95 \pm 0.5$ \\
$\mathcal{M}^{\texttt{SVM}}_
\mathit{hashtag}$       & $42.69 \pm 0.03$ & $100.00 \pm 0$ & $59.83 \pm 0.03$ \\
\hline
\end{tabular}%
}
\vspace{0.5cm}
\caption{Test performance comparison. Five runs per experiment were conducted and mean and standard deviation are presented. }
\label{tab:train_result}
\end{table}



\section{Related Work}

Social media response to the ongoing pandemic has received significant research attention: (1) health misinformation ~\cite{CarleyMisinformation,hossain2020covidlies,cinelli2020covid}, (2) polarization ~\cite{CarleyCovidPolarization,khudabukhsh2020}, (3) disease modeling~\cite{li2020retrospective}, etc. Counterhate measures along the line of counterspeech research~\cite{benesch2016counterspeech, benesch2014defining, mathew2018analyzing,AAAIRohingya} to combat Anti-Asian hate~\cite{DBLP:journals/corr/abs-2005-12423}, and community blame~\cite{FearSpeech} has been studied. Our work contrasts with existing literature in three ways: (1) we analyze bilateral relations of nuclear adversaries amidst a raging pandemic; (2) we release a novel data set for wider use exploring related research questions; and (3) we present a new method that combines recent \emph{NLP for positive impact} advances in a new, timely, and important task.

While the political volatility between India and Pakistan has been extensively studied by social scientists~\cite{malik2002kashmir, schofield2010kashmir, bose2009kashmir}, barring few recent lines of work~\cite{ECAIHopeSpeech,KhudaBukhshHarnessing, TyagiSocInfo}, social media interactions between the civilians of India and Pakistan has received little or no attention. All recent work on Indian and Pakistani social media ~\cite{ECAIHopeSpeech,KhudaBukhshHarnessing, TyagiSocInfo} focused on a solitary incident - the 2019 India-Pakistan conflict triggered by the Pulwama terror attack across different social media platforms. While~\cite{ECAIHopeSpeech} introduced a novel task of detecting hostility-diffusing, peace seeking \emph{hope speech} and considered comments on relevant YouTube videos as the data set,~\cite{TyagiSocInfo} is the first work on analyzing web-manifestation (Twitter) of political polarization between the two countries and how political parties factor in these discussions.

Our work leverages two existing resources: (1) a \emph{hope speech} classifier introduced in~\cite{ECAIHopeSpeech}; and (2) a well-known \emph{empathy-distress} data set~\cite{buechel-etal-2018-modeling}. As already mentioned, our work differs in a key way that we re-purpose these resources for a new \emph{NLP for positive impact} task: detecting \emph{supportive} tweets in the context of social media discussions during a national healthcare crisis. Our work also draws inspiration from recent findings about mining stance from hashtags~\cite{Kumar2018WeaklySS}.


\section{Methods, Results, and Discussion}
\noindent\textbf{Research question}: \emph{Does sampling tweets containing $\mathcal{H}_\mathit{supportive}$ hashtags alone suffice?}

We first investigate if hashtag-based filtering alone guarantees \emph{supportive} tweets with a high probability. We randomly sample 1,000 tweet texts from $\mathcal{T}_\mathit{supportive}$ and manually annotate them. Our annotators are provided only the tweet texts, i.e., the body of the tweet excluding hashtags, urls, mentions, and emojis. Three annotators fluent in English, Hindi, and Urdu, and well-versed with the geopolitical events between India and Pakistan first independently annotated these tweets and achieved a Fleiss' $\kappa$ score of 0.76 indicating moderate agreement. Next, disagreements are resolved through a follow-up adjudication process and a higher Fleiss' $\kappa$ score of 0.86 is reached. Of the randomly chosen 1,000 tweets 444 tweets, i.e., 44.4\% were marked positive. This result indicates that solely relying on \emph{supportive} hashtag will not do better than chance and underscores the importance of sophisticated methods. 

In addition, we randomly sampled  1,000 tweet texts from $\mathcal{T}_\mathit{supportive} \cup \mathcal{T}_{\mathit{not}\emph{-}\mathit{supportive}}$ as our test set (denoted as $\mathcal{D}_\mathit{eval}$). Throughout our annotation process, whenever consensus label is absent, following standard literature~\cite{manningEntailment}, we consider the majority label as the gold-standard label. Annotator subjectivity is a well-studied research area~\cite{pavlick2019inherent}, and in order to facilitate further research, we also provide individual annotator's labels.


\noindent\textbf{Research question}: \emph{Do the hope speech and the empathy-distress classifiers present any discernible signal to differentiate between supportive and not-supportive tweets?}

As already described, the \emph{hope speech} classifier is designed for a different scenario of detecting peace-seeking, hostility diffusing content from social media discussions generated during a conflict. Our current task of detecting \emph{supportive} tweets, although related, is not identical. Furthermore, the classifier is trained on a different social media platform, YouTube, that allows unstructured text without any length restriction, whereas Twitter allows unstructured text but imposes a length restriction. Similarly, the \emph{empathy-distress} classifier is trained on a different data set of user responses to news events. Hence, a pertinent research question is if the \emph{hope speech} classifier or the \emph{empathy-distress} classifier is any good in differentiating between \emph{supportive} and \emph{not-supportive} tweets. 

We first start with a simple experiment to illustrate that the resources provide useful signal. Let $S$ =  $\{\langle x, y \rangle\}$ such that $x \sim  \mathcal{T}_{\mathit{supportive}}$ and $y \sim \mathcal{T}_{ \mathit{not}\emph{-}\mathit{supportive}}$, i.e., $S$ consists of tweet pairs $\langle x, y \rangle$ where $x$ and $y$ are randomly drawn from the pool of tweets with \emph{supportive} and \emph{not-supportive} hashtags, respectively. Let $\mathcal{P}_{h} (z)$ and $\mathcal{P}_{e} (z)$ denote the predicted \emph{hope speech} and \emph{empathy-distress} probabilities of tweet $z$. We compute:\\ $r_h = \frac{\Sigma_{\langle x, y \rangle \in \mathcal{S}_\mathit{s}}\mathbbm{I}(\mathcal{P}_{h} (x) > \mathcal{P}_{h} (y) )}{|\mathcal{S}|}$ and \\
$r_e = \frac{\Sigma_{\langle x, y \rangle \in \mathcal{S}_\mathit{s}}\mathbbm{I}(\mathcal{P}_{e} (x) > \mathcal{P}_{e} (y) )}{|\mathcal{S}|}$
where $\mathbbm{I}$ denotes an indicator function and $|\mathcal{S}|$, i.e., the number of randomly drawn pairs, is set to 100,000. We ran this experiment five times and found $r_h$ to be equal to  69.3 $\pm$ 0.13\% and $r_e$ to be equal to   47.8 $\pm$ 0.12\%,  indicating that a randomly drawn sample from $\mathcal{T}_\mathit{supportive}$ is more likely to receive a higher \emph{hope speech} score ($\mathcal{P}_{h} (.)$) than a randomly drawn sample from $\mathcal{T}_{ \mathit{not}\emph{-}\mathit{supportive}}$. However, we do not notice similar trends with our \emph{empathy-distress} classifier.  

It is unsurprising that  $r_h$ has a much higher value than $r_e$. The \emph{hope speech} classifier is trained on a data set relevant to a recent India-Pakistan conflict and thus has a substantial overlap in domain. Hence, a general nature of positive dialogue may indicate a desire to put things behind and help each other. In contrast, the \emph{empathy-distress} classifier is trained on a broad, diverse, data set of user responses to news events and has no overlap with the current domain. However, when we rank tweets from $\mathcal{T}_\mathit{supportive}$ by the classifier's probability, we notice that top predictions are of extremely high quality in both cases. We annotate top 1,000 unique tweets from $\mathcal{T}_\mathit{supportive}$ ranked by $\mathcal{P}_{h} (.)$ and obtain 950 positives. Similarly, top 1,000 unique tweets from $\mathcal{T}_\mathit{supportive}$ ranked by $\mathcal{P}_{e} (.)$ yield 899 positives upon manual annotation. Moreover, the two classifiers complement each other as among the top 1,000 unique tweets from the \emph{hope speech} classifier and the top 1,000 unique tweets from the \emph{empathy-distress} classifier had minimal overlap (62 samples). This annotation task also yielded a substantially higher Fleiss' $\kappa$ score ($0.8068$) without any follow-up adjudication process indicating that the chosen samples have less ambiguity than our earlier experiment that involved annotating randomly selected tweets from $\mathcal{T}_\mathit{supportive}$. Our results thus indicate existing resources can be harnessed for informed sampling yielding high-quality positives.  

\begin{algorithm*}[htb]
\small
\DontPrintSemicolon
\SetAlgoLined

\textbf{Input:} $\mathcal{T}$ is the full set of tweets, $\mathcal{T}_\mathit{supportive}$, $\mathcal{T}_{\mathit{not}\emph{-}\mathit{supportive}}$ $\subset \mathcal{T}$;  $\mathcal{M}_\mathit{hopeSpeech}$ is the hope speech classifier; $\mathcal{M}_\mathit{empathy Distress}$ is the empathy-distress classifier\\
\textbf{Output:} $\mathcal{D}_\mathit{informed} \subset \mathcal{T}$; and  $\mathcal{M}_\mathit{informed}$ - a model trained on $\mathcal{D}_\mathit{informed}$\\

\textbf{Procedure:}\;
\ForEach{\emph{tweet} ~t~$\in$~$\mathcal{T}_\mathit{supportive} \cup \mathcal{T}_{\mathit{not}\emph{-}\mathit{supportive}}$ }
{
classify $t$ using $\mathcal{M}_\mathit{hopeSpeech}$ and $\mathcal{M}_\mathit{empathyDistress}$ yielding positive probabilities $\mathcal{P}_h$ and $\mathcal{P}_e$. 

}

Sort $\mathcal{T}_\mathit{supportive}$ using $\mathcal{P}_h$ and $\mathcal{P}_e$ yielding two ranked lists $\mathcal{R}_{\mathit{supportive}_h}$ and $\mathcal{R}_{\mathit{supportive}_e}$.

Take the top 1,000 tweets from $\mathcal{R}_{\mathit{supportive}_h}$ and $\mathcal{R}_{\mathit{supportive}_e}$ yielding 2,000 tweets - these are the positive samples - ${\mathcal{D}^{+}}_\mathit{informed}$.

Sort $\mathcal{T}_{\mathit{not}\emph{-}\mathit{supportive}}$ using $\mathcal{P}_h$ and $\mathcal{P}_e$ yielding two ranked lists $\mathcal{R}_{{\mathit{not}\emph{-}\mathit{supportive}}_h}$ and $\mathcal{R}_{{\mathit{not}\emph{-}\mathit{supportive}}_e}$.

Sample 500 tweets from the bottom 80\% of $\mathcal{R}_h$ and $\mathcal{R}_e$ yielding 1,000 tweets - these are the negative samples - ${\mathcal{D}^{-}}_\mathit{informed}$.

$\mathcal{D}_\mathit{informed} \leftarrow {\mathcal{D}^{+}}_\mathit{informed} \cup {\mathcal{D}^{-}}_{informed}$\\

Duplicates are discarded from $\mathcal{D}_\mathit{informed}$

$\mathcal{M}_\mathit{informed} \leftarrow $ a classifier trained on $\mathcal{D}_\mathit{informed}$\\

\textbf{Output:} $\mathcal{D}_\mathit{informed}$ and $\mathcal{M}_\mathit{informed}$
\caption{{ {\small{$\mathit{Construct}$($\mathcal{D}_\mathit{informed}$, $\mathcal{M}_\mathit{informed}$)}}}}

\label{algo:NN-algo}

\end{algorithm*}

\noindent\textbf{Research question}: \emph{How to leverage existing resources to design an effective classifier to detect supportive tweets?}

We utilize two existing resources, a \emph{hope speech} classifier from~\cite{ECAIHopeSpeech}, and an \emph{empathy-distress} data set from~\cite{buechel-etal-2018-modeling}. We first train an \emph{empathy-distress} classifier on the \emph{empathy-distress} data set that can classify tweets as exhibiting empathy or distress, or not.


Our pipeline utilizes the \emph{hope speech} and \emph{empathy-distress} classifiers and constructs a weakly labeled data set where the positive examples exhibit themes like empathy, distress, support, and solidarity - the \emph{supportive speech}, and the negative examples exhibit themes like controversy, whataboutism, and hostility - the \emph{not-supportive speech}. The two classifiers are used to label tweets and the positive class probability is used to rank all the tweets in the set $\mathcal{T}_{\mathit{supportive}} \cup \mathcal{T}_{\mathit{not}\emph{-}\mathit{supportive}}$ yielding two ranked lists.     ${\mathcal{D}^{+}}_{\mathit{informed}}$ contains all tweets using any of the top 1,000 tweets in both ranked lists (2,000 in total, 1,938 unique) are considered positive samples, and a set of negative samples, ${\mathcal{D}^{-}}_{\mathit{informed}}$, is constructed by randomly sampling 500 tweets each from the bottom 80\% of both ranked lists (1,000 in total, 1,000 unique). The full data set construction pipeline is presented in Algorithm~\ref{algo:NN-algo}. The trained model is denoted as $\mathcal{M}_\mathit{informed}$. Table~\ref{tab:dinformed} lists a random sample of tweet texts from $\mathcal{D}_\mathit{informed}$.

\begin{table}[htb]
{
\small
\begin{center}
     \begin{tabular}{|p{0.80\textwidth}|}
   
   \hline
Prayers for India we are with you May Allah Almighty protect all Indians from this deadly virus Ameen \\
\hline 
We have boundries but not in our hearts\\
We are humans, we have pain\\
We are Neighbours not Enemies\\
Humanity First.\\
Prayers for India \\
   \hline 
It doesn't matter how many differences there are between our countries. But humanity first.
we all are in it together. I hope soon thing comes in control IA.our prayers with the people’s of India get well soon neighbors \\
    \hline
I request our government to extend the hands to help people of India in this difficult time.\\
May Allah ease the pain of our neighbour. Horrible situation in india as country Just ran out of oxygen\\
Prayers and greeting from pakistan.
\\
\hline 
Heartbreaking to see this situation in our neighbourhood.\\
Send love and prayers from Pakistan. May Almighty Allah help humanity through this pandemic.\\
Stay strong, Stay Safe
\\
\hline
    \end{tabular}
\end{center}
\vspace{0.5cm}
\caption{{Randomly sampled tweet texts from $\mathcal{D}^{+}_\mathit{informed}$ annotated as positives.}}  
\label{tab:dinformed}}
\end{table}

Earlier research has reported hashtags as an effective way to obtain weak labels~\cite{Kumar2018WeaklySS}. 
We contrast $\mathcal{M}_\mathit{informed}$ against a baseline that uses hashtags alone as a source of weak labels and contains the identical number of (weakly labeled) positives and negatives as $\mathcal{D}_\mathit{informed}$. Essentially, any tweet belonging to $\mathcal{T}_{\mathit{supportive}}$ is considered a positive and any tweet belonging to $\mathcal{T}_{\mathit{not}\emph{-}\mathit{supportive}}$ is considered a negative. Positives and negative examples are randomly sampled from these sets and a data set with the same proportions as $\mathcal{D}_\mathit{informed}$ is constructed. The trained model is denoted as $\mathcal{M}_\mathit{hashtag}$.

We train our classifiers using  \texttt{BERT}~\cite{Devlin2019BERTPO} (\texttt{bert-base-uncased}) using the transformers library~\cite{wolf-etal-2020-transformers} and a 90/10 train/validation split. In addition, since English social media content from the Indian subcontinent exhibits a variety of disfluencies~\cite{sarkar-etal-2020-non}, and since the SVM baseline has been successfully applied to the original \emph{hope speech} detection task~\cite{ECAIHopeSpeech}, we include an SVM baseline as well that uses TF-IDF vectors as document feature representations. The trained models are evaluated on $\mathcal{D}_\mathit{eval}$, 1000 randomly sampled tweets from $\mathcal{T}_{\mathit{supportive}} \cup \mathcal{T}_{\mathit{not}\emph{-}\mathit{supportive}}$.  Note that hashtags, urls, emojis, mentions, and punctuation are removed from the tweets prior to training.

\subsection{Performance Comparison}

\begin{table}[htb]
{
\small
\begin{center}
     \begin{tabular}{|p{0.80\textwidth}|}
   
   \hline
Life is dying in our neighboring country. We have differences. We have fought wars, but we are neighbors. Sighing lives in India.  My lord, who will do good except you\\ There is no religion of humanity. May Allah save the whole world including India from this epidemic. Amen \\
\hline 
From Pakistan I request my all Muslims \\ Humanity has no religion and no boundaries ....Pray for all  world and for India \\
   \hline 
Be safe everone, wear mask everytime, may your country doesn't goes through what our country is going. Greetings from india \\
    \hline
    \end{tabular}
\end{center}
\caption{{Randomly sampled YouTube comments predicted as \emph{supportive} by $\mathcal{M}^{\texttt{BERT}}_\mathit{informed}$ in the wild.}}  
\label{tab:inthewild}}
\end{table}

Table~\ref{tab:train_result} shows that $\mathcal{M}_\mathit{informed}$ substantially outperforms $\mathcal{M}_\mathit{hashtag}$ on the test set and thus underscores why hashtag-based-filtering may not solely suffice. Also, this result indicates that the joint concept of empathy, distress, and solidarity is learnable, and in this context, the resources exhibit synergy. Understandably, a supervised solution will improve the performance since weak labels obtained using the \emph{hope speech} and \emph{empathy-distress} classifier, while high-quality, still had some amount of noise. Compared to the informed sampling, we observe a slight performance boost in our supervised solutions. We also notice the \texttt{BERT}-based classifiers outperformed SVM baselines.  

While our primary focus is on Twitter, several social media platforms exist where hashtags are not as prevalent. YouTube, a highly popular social media platform, is one such example. We performed an in-the-wild test where we obtained the top 100 \emph{supportive} predictions from a new data set consisting of 31,232 comments on 185 YouTube COVID-19-related videos from the official YouTube channel of Geo TV, a highly popular Pakistani news channel. We used the best $\mathcal{M}^\texttt{BERT}_{informed}$ model to test our minimally supervised method's in-the-wild performance. Out of 100 such comments, a manual evaluation revealed that 70 were positive.  Table~\ref{tab:inthewild} lists a few such randomly sampled comments. A reasonably high precision of our model indicates its cross-platform viability and applicability in downstream tasks like moderation. 

\subsection{Discussion}
\noindent\textbf{Research question:} \emph{How Pakistan Responded to this crisis?} In our earlier analysis in Section~\ref{sec:likeRetweet}, we found that tweets containing $\mathcal{H}_{\mathit{supportive}}$ hashtags originating in Pakistan (1) heavily outnumbered those containing $\mathcal{H}_{\mathit{not}\emph{-}\mathit{supportive}}$ hashtags; and (2) received a larger share of the likes and retweets. We investigate the like and retweet behavior conditioned on the tweet text less the hashtags. Table~\ref{tab:likeRetweetFromPakistan} indicates an overwhelming majority of the tweets from Pakistan is classified as \emph{supportive} by $\mathcal{M}_\mathit{supervised}$ and such tweets received substantially more likes and retweets than the \emph{not-supportive} tweets.

 \begin{table}[htb]
\centering
\small
\resizebox{0.60\textwidth}{!}{%
\centering
\begin{tabular}{llll}
\hline
Label & Percentage & Like   & Retweet \\ \hline
supportive$_\mathit{Pak}$ & 85.30\% & $6.64 \pm 270.6$ &  $505.61 \pm 1378.1$ \\
not-supportive$_\mathit{Pak}$ & 14.70\% & $1.26 \pm 24.8$ & $276.28 \pm 300.9$  \\
\hline
\end{tabular}%
}
\vspace{0.5cm}
\caption{Like and retweet behavior and count of \emph{supportive} and \emph{not-supportive} tweets from Pakistan.}

\label{tab:likeRetweetFromPakistan}
\end{table}

\section{Ethical and Societal Implications}

While the setting discussed in the paper involves  humanitarian tasks, the techniques can be trivially adapted with the explicit objective to censor empathetic content. In many recent conflicts in the Indian subcontinent, such systems can have adverse social effects, and thus particular care is needed before these systems are deployed. Also, language-specific features can sometimes cause syntactically similar but semantically opposite content to be surfaced underscoring the need for a human-in-the-loop setting before such systems are deployed for social media content moderation tasks. Finally, our classifier relies on a black box \emph{hope speech} classifier and thus runs the risk of propagating possible biases from the black box model. Further case studies need to be considered before deployment and we welcome a thorough investigation of our released data set from the scientific community.   

\section{Conclusions}

In this paper, we present a task and associated resources for a vital domain - geopolitical relations against the backdrop of a raging pandemic. We release a data set of tweets discussing the oxygen crisis and healthcare system collapse in India due to a COVID-19 wave. Our data set is geographically diverse and connects several diverse themes - a long acrimonious history between two neighboring countries that involves four wars and a recent bilateral relations breakdown, a raging pandemic that has claimed several hundred thousand lives within a few weeks and is still ongoing. Our analysis reveals a strong humanitarian streak that prioritizes health and well-being over past geographical or ethnic disputes. We then re-purpose existing resources designed for adjacent tasks like \emph{hope speech} and \emph{empathy distress} detection and utilize these to identify \emph{supportive} tweets. Our experiments reveal that \emph{NLP for positive impact} tasks can utilize existing adjacent resources to rapidly bootstrap solutions.

\bibliographystyle{unsrt}

\begin{thebibliography}{10}

\bibitem{carvalho2021first}
Thiago Carvalho, Florian Krammer, and Akiko Iwasaki.
\newblock The first 12 months of covid-19: a timeline of immunological
  insights.
\newblock {\em Nature Reviews Immunology}, 21(4):245--256, 2021.

\bibitem{CNN}
India is spiraling deeper into covid-19 crisis. here's what you need to know.
\newblock
  \url{https://www.cnn.com/2021/04/26/india/india-covid-second-wave-explainer-intl-hnk-dst/index.html},
  2021.
\newblock Online; accessed 7-June-2021.

\bibitem{BBC}
Covid: India sees world's highest daily cases amid oxygen shortage.
\newblock \url{https://www.bbc.com/news/world-asia-india-56826645}, 2021.
\newblock Online; accessed 7-June-2021.

\bibitem{paul2005india}
Thazha~Varkey Paul and Thazha~Varkey Paul.
\newblock {\em The India-Pakistan conflict: an enduring rivalry}.
\newblock Cambridge University Press, 2005.

\bibitem{ECAIHopeSpeech}
Shriphani Palakodety, Ashiqur~R. KhudaBukhsh, and Jaime~G. Carbonell.
\newblock Hope speech detection: {A} computational analysis of the voice of
  peace.
\newblock In Giuseppe~De Giacomo, Alejandro Catal{\'{a}}, Bistra Dilkina,
  Michela Milano, Sen{\'{e}}n Barro, Alberto Bugar{\'{\i}}n, and
  J{\'{e}}r{\^{o}}me Lang, editors, {\em {ECAI} 2020 - 24th European Conference
  on Artificial Intelligence}, volume 325 of {\em Frontiers in Artificial
  Intelligence and Applications}, pages 1881--1889. {IOS} Press, 2020.

\bibitem{buechel-etal-2018-modeling}
Sven Buechel, Anneke Buffone, Barry Slaff, Lyle Ungar, and Jo{\~a}o Sedoc.
\newblock Modeling empathy and distress in reaction to news stories.
\newblock In {\em Proceedings of the 2018 Conference on Empirical Methods in
  Natural Language Processing}, pages 4758--4765, Brussels, Belgium,
  October-November 2018. Association for Computational Linguistics.

\bibitem{arango2019hate}
Aym{\'e} Arango, Jorge P{\'e}rez, and Barbara Poblete.
\newblock Hate speech detection is not as easy as you may think: A closer look
  at model validation.
\newblock In {\em Proceedings of the 42nd international acm sigir conference on
  research and development in information retrieval}, pages 45--54, 2019.

\bibitem{Beltagy2019SciBERT}
Iz~Beltagy, Kyle Lo, and Arman Cohan.
\newblock Scibert: Pretrained language model for scientific text.
\newblock In {\em EMNLP}, 2019.

\bibitem{TyagiSocInfo}
Aman Tyagi, Anjalie Field, Priyank Lathwal, Yulia Tsvetkov, and Kathleen~M.
  Carley.
\newblock A computational analysis of polarization on indian and pakistani
  social media.
\newblock In Samin Aref, Kalina Bontcheva, Marco Braghieri, Frank Dignum, Fosca
  Giannotti, Francesco Grisolia, and Dino Pedreschi, editors, {\em Social
  Informatics - 12th International Conference, SocInfo 2020, Pisa, Italy,
  October 6-9, 2020, Proceedings}, volume 12467 of {\em Lecture Notes in
  Computer Science}, pages 364--379. Springer, 2020.

\bibitem{chhaya2020editorial}
Niyati Chhaya, Kokil Jaidka, Lyle Ungar, Jennifer Healey, and Atanu Sinha.
\newblock Editorial for the 3rd aaai-20 workshop on affective content analysis.
\newblock 2020.

\bibitem{batson1987distress}
C~Daniel Batson, Jim Fultz, and Patricia~A Schoenrade.
\newblock Distress and empathy: Two qualitatively distinct vicarious emotions
  with different motivational consequences.
\newblock {\em Journal of personality}, 55(1):19--39, 1987.

\bibitem{batson1991evidence}
C~Daniel Batson and Laura~L Shaw.
\newblock Evidence for altruism: Toward a pluralism of prosocial motives.
\newblock {\em Psychological inquiry}, 2(2):107--122, 1991.

\bibitem{sober1999unto}
Elliot Sober and David~Sloan Wilson.
\newblock {\em Unto others: The evolution and psychology of unselfish
  behavior}.
\newblock Number 218. Harvard University Press, 1999.

\bibitem{goetz2010compassion}
Jennifer~L Goetz, Dacher Keltner, and Emiliana Simon-Thomas.
\newblock Compassion: an evolutionary analysis and empirical review.
\newblock {\em Psychological bulletin}, 136(3):351, 2010.

\bibitem{mikulincer2010prosocial}
Mario~Ed Mikulincer and Phillip~R Shaver.
\newblock {\em Prosocial motives, emotions, and behavior: The better angels of
  our nature.}
\newblock American Psychological Association, 2010.

\bibitem{wildt1999solidarity}
Andreas Wildt.
\newblock Solidarity: its history and contemporary definition.
\newblock In {\em Solidarity}, pages 209--220. Springer, 1999.

\bibitem{Devlin2019BERTPO}
J.~Devlin, Ming-Wei Chang, Kenton Lee, and Kristina Toutanova.
\newblock Bert: Pre-training of deep bidirectional transformers for language
  understanding.
\newblock In {\em NAACL-HLT}, 2019.

\bibitem{wolf-etal-2020-transformers}
Thomas Wolf, Lysandre Debut, Victor Sanh, Julien Chaumond, Clement Delangue,
  Anthony Moi, Pierric Cistac, Tim Rault, Rémi Louf, Morgan Funtowicz, Joe
  Davison, Sam Shleifer, Patrick von Platen, Clara Ma, Yacine Jernite, Julien
  Plu, Canwen Xu, Teven~Le Scao, Sylvain Gugger, Mariama Drame, Quentin Lhoest,
  and Alexander~M. Rush.
\newblock Transformers: State-of-the-art natural language processing.
\newblock In {\em Proceedings of the 2020 Conference on Empirical Methods in
  Natural Language Processing: System Demonstrations}, pages 38--45, Online,
  October 2020. Association for Computational Linguistics.

\bibitem{sindhwani2009uncertainty}
Vikas Sindhwani, Prem Melville, and Richard~D Lawrence.
\newblock Uncertainty sampling and transductive experimental design for active
  dual supervision.
\newblock In {\em Proceedings of the 26th Annual International Conference on
  Machine Learning}, pages 953--960, 2009.

\bibitem{attenberg2010unified}
Josh Attenberg, Prem Melville, and Foster Provost.
\newblock A unified approach to active dual supervision for labeling features
  and examples.
\newblock In {\em Joint European Conference on Machine Learning and Knowledge
  Discovery in Databases}, pages 40--55. Springer, 2010.

\bibitem{khudabukhsh2015building}
Ashiqur~R KhudaBukhsh, Paul~N Bennett, and Ryen~W White.
\newblock Building effective query classifiers: a case study in self-harm
  intent detection.
\newblock In {\em Proceedings of the 24th ACM International on Conference on
  Information and Knowledge Management}, pages 1735--1738, 2015.

\bibitem{rudra-etal-2016-understanding}
Koustav Rudra, Shruti Rijhwani, Rafiya Begum, Kalika Bali, Monojit Choudhury,
  and Niloy Ganguly.
\newblock Understanding language preference for expression of opinion and
  sentiment: What do {H}indi-{E}nglish speakers do on {T}witter?
\newblock In {\em Proceedings of the 2016 Conference on Empirical Methods in
  Natural Language Processing}, pages 1131--1141, Austin, Texas, November 2016.
  Association for Computational Linguistics.

\bibitem{KhudaBukhshHarnessing}
Ashiqur~R. KhudaBukhsh, Shriphani Palakodety, and Jaime~G. Carbonell.
\newblock Harnessing code switching to transcend the linguistic barrier.
\newblock In Christian Bessiere, editor, {\em Proceedings of the Twenty-Ninth
  International Joint Conference on Artificial Intelligence, {IJCAI} 2020},
  pages 4366--4374. ijcai.org, 2020.

\bibitem{CarleyMisinformation}
Shahan~Ali Memon and Kathleen~M. Carley.
\newblock Characterizing {COVID-19} misinformation communities using a novel
  twitter dataset.
\newblock In Stefan Conrad and Ilaria Tiddi, editors, {\em Proceedings of the
  {CIKM} 2020 Workshops co-located with 29th {ACM} International Conference on
  Information and Knowledge Management {(CIKM} 2020), Galway, Ireland, October
  19-23, 2020}, volume 2699 of {\em {CEUR} Workshop Proceedings}. CEUR-WS.org,
  2020.

\bibitem{hossain2020covidlies}
Tamanna Hossain, Robert~L Logan~IV, Arjuna Ugarte, Yoshitomo Matsubara, Sean
  Young, and Sameer Singh.
\newblock Covidlies: Detecting covid-19 misinformation on social media.
\newblock In {\em Proceedings of the 1st Workshop on NLP for COVID-19 (Part 2)
  at EMNLP 2020}, 2020.

\bibitem{cinelli2020covid}
Matteo Cinelli, Walter Quattrociocchi, Alessandro Galeazzi, Carlo~Michele
  Valensise, Emanuele Brugnoli, Ana~Lucia Schmidt, Paola Zola, Fabiana Zollo,
  and Antonio Scala.
\newblock The covid-19 social media infodemic.
\newblock {\em Scientific Reports}, 10(1):1--10, 2020.

\bibitem{CarleyCovidPolarization}
Iain~J. Cruickshank and Kathleen~M. Carley.
\newblock Characterizing communities of hashtag usage on twitter during the
  2020 {COVID-19} pandemic by multi-view clustering.
\newblock {\em Appl. Netw. Sci.}, 5(1):66, 2020.

\bibitem{khudabukhsh2020}
Ashiqur~R. KhudaBukhsh, Rupak Sarkar, Mark~S. Kamlet, and Tom~M. Mitchell.
\newblock We don't speak the same language: Interpreting polarization through
  machine translation.
\newblock In {\em The Thirty-Fifth AAAI Conference on Artificial Intelligence,
  {AAAI} 2021}, page To Appear. {AAAI} Press, 2021.

\bibitem{li2020retrospective}
Cuilian Li, Li~Jia Chen, Xueyu Chen, Mingzhi Zhang, Chi~Pui Pang, and Haoyu
  Chen.
\newblock Retrospective analysis of the possibility of predicting the covid-19
  outbreak from internet searches and social media data, china, 2020.
\newblock {\em Eurosurveillance}, 25(10):2000199, 2020.

\bibitem{benesch2016counterspeech}
Susan Benesch, Derek Ruths, Kelly~P Dillon, Haji~Mohammad Saleem, and Lucas
  Wright.
\newblock Counterspeech on twitter: A field study.
\newblock {\em A report for Public Safety Canada under the Kanishka Project},
  2016.

\bibitem{benesch2014defining}
Susan Benesch.
\newblock Defining and diminishing hate speech.
\newblock {\em State of the World’s Minorities and Indigenous Peoples},
  2014:18--25, 2014.

\bibitem{mathew2018analyzing}
Binny Mathew, Navish Kumar, Pawan Goyal, Animesh Mukherjee, et~al.
\newblock Analyzing the hate and counter speech accounts on twitter.
\newblock {\em arXiv preprint arXiv:1812.02712}, 2018.

\bibitem{AAAIRohingya}
Shriphani Palakodety, Ashiqur~R. KhudaBukhsh, and Jaime~G. Carbonell.
\newblock Voice for the voiceless: Active sampling to detect comments
  supporting the rohingyas.
\newblock In {\em The Thirty-Fourth {AAAI} Conference on Artificial
  Intelligence}, pages 454--462. {AAAI} Press, 2020.

\bibitem{DBLP:journals/corr/abs-2005-12423}
Caleb Ziems, Bing He, Sandeep Soni, and Srijan Kumar.
\newblock Racism is a virus: Anti-asian hate and counterhate in social media
  during the {COVID-19} crisis.
\newblock {\em CoRR}, abs/2005.12423, 2020.

\bibitem{FearSpeech}
Punyajoy Saha, Binny Mathew, Kiran Garimella, and Animesh Mukherjee.
\newblock "short is the road that leads from fear to hate": Fear speech in
  indian whatsapp groups.
\newblock {\em CoRR}, abs/2102.03870, 2021.

\bibitem{malik2002kashmir}
Iffat Malik and Robert~G Wirsing.
\newblock {\em Kashmir: Ethnic conflict international dispute}.
\newblock Oxford University Press Oxford, 2002.

\bibitem{schofield2010kashmir}
Victoria Schofield.
\newblock {\em Kashmir in conflict: India, Pakistan and the unending war}.
\newblock Bloomsbury Publishing, 2010.

\bibitem{bose2009kashmir}
Sumantra Bose.
\newblock {\em Kashmir: Roots of conflict, paths to peace}.
\newblock Harvard University Press, 2009.

\bibitem{Kumar2018WeaklySS}
Sumeet Kumar.
\newblock Weakly supervised stance learning using social-media hashtags.
\newblock 2018.

\bibitem{manningEntailment}
Samuel~R. Bowman, Gabor Angeli, Christopher Potts, and Christopher~D. Manning.
\newblock A large annotated corpus for learning natural language inference.
\newblock In Llu{\'{\i}}s M{\`{a}}rquez, Chris Callison{-}Burch, Jian Su,
  Daniele Pighin, and Yuval Marton, editors, {\em Proceedings of the 2015
  Conference on Empirical Methods in Natural Language Processing, {EMNLP} 2015,
  Lisbon, Portugal, September 17-21, 2015}, pages 632--642. The Association for
  Computational Linguistics, 2015.

\bibitem{pavlick2019inherent}
Ellie Pavlick and Tom Kwiatkowski.
\newblock Inherent disagreements in human textual inferences.
\newblock {\em Transactions of the Association for Computational Linguistics},
  7:677--694, 2019.

\bibitem{sarkar-etal-2020-non}
Rupak Sarkar, Sayantan Mahinder, and Ashiqur KhudaBukhsh.
\newblock The non-native speaker aspect: {I}ndian {E}nglish in social media.
\newblock In {\em Proceedings of the Sixth Workshop on Noisy User-generated
  Text (W-NUT 2020)}, pages 61--70, Online, November 2020. Association for
  Computational Linguistics.

\end{thebibliography}


\end{document}